\documentclass[sigplan,screen]{acmart}
\AtBeginDocument{%
  \providecommand\BibTeX{{%
    \normalfont B\kern-0.5em{\scshape i\kern-0.25em b}\kern-0.8em\TeX}}}
\begin{document}
\title{Unsupervised Identification of Relevant Prior Cases}
\author{Shivangi Bithel}
\email{cs20mtech12004@iith.ac.in}
\affiliation{%
  \institution{Indian Institute of Technology}
  \city{Hyderabad}
  \state{Telangana}
  \country{India}
}

\author{Sumitra S Malagi}
\email{cs20mtech14006@iith.ac.in}
\affiliation{%
  \institution{Indian Institute of Technology}
  \city{Hyderabad}
  \state{Telangana}
  \country{India}
}
\begin{abstract}
Document retrieval has taken its role in almost all domains of knowledge understanding including the legal domain. Precedent refers to a court decision that is considered as authority for deciding subsequent cases involving identical or similar facts, or similar legal issues. In this work, we are proposing different unsupervised approaches to solve the task of identifying relevant precedents to a given query case. Our proposed approaches are using word embeddings like word2vec, doc2vec and sent2vec, finding cosine similarity using TF-IDF, retrieving relevant documents using BM25 scores, using pretrained model and SBERT to find the most similar document and using product of BM25 and TF-IDF scores to find the most relevant document for a given query. We compared all the methods based on precision@10, recall@10 and MRR. Based on the comparative analysis we found that TF-IDF score multiplied with BM25 score gives the best result. In this paper we have also presented the analysis that we did to improve the BM25 score.
\end{abstract}
\maketitle
\section{Introduction}
Legal practitioners working on new legal case refer to previously judged similar cases. This helps them understand how the proceeding for similar cases was held and how it was judged. With humongous number of prior cases to refer from, it’s very cumbersome to manually select the most relevant case document, hence automating the task of retrieving relevant prior cases is advantageous. This kind of automated solution is not only helpful for lawyers, but also to common man who would like to understand how cases like theirs were dealt in past and they can approach the lawyer accordingly. It will assist the user in identifying under which court (i.e criminal/civil etc) their case fits, how they can proceed with the case and what were the outcomes of similar cases. The goal is given a query, i.e description of a case, we build an information retrieval system that will rank documents in the corpus based on their relevancy. The corpus consists of 2914 prior cases. We were given 50 queries; these queries were summarized case descriptions. We tried numerous methods to build our system, all the approaches we used were unsupervised as we had very less data and the data is unbalanced. The rest of this paper is organized as follows: Section 2 presents related work done that have tried to solve information retrieval task for legal and also in other domains. Section 3 provides the task description and summarizes about dataset. In Section 4 and 5, we present how data pre-processing was done and the various approaches we used. In section 6 and 7, we will brief about the results and various analysis that we carried out. Section 8 concludes the work and suggest some future improvements that can be done.
\section{Related Work}
Since the task of retrieving relevant documents is very helpful for lawyers across the globe, there has been considerable work in this direction. Task1 of AILA@FIRE is also about retrieving relevant prior case for a given query. In one of the past submissions \cite{article1} of AILA, participants have computed TF-IDF vectors of query and case documents. Prior cases were retrieved based on cosine similarity with respect to the query. Another team \cite{article2} that participated in AILA@2019 initially used word2vec to find word embeddings and later improvised the embedding with Glove, but the results were not as expected, as Glove and word2vec models are not trained on legal data. Similar task was introduced in Competition on Legal Information Extraction and Entailment (COLIEE-2019), as an associated event of International Conference on Artificial Intelligence and Law (ICAIL)-2019, one of the participating teams \cite{article3} has used doc2vec similarity, later they calculated the BM25 score with respect to query and multiplied it with doc2vec similarity. We took inspiration from them and calculated a combined score using cosine similarity obtained from TF-IDF vectors and BM25 similarity scores. There has also been significant work in ad-hoc document retrieval, Canjia Li \cite{article4}  et al. have presented PARADE, an end-to-end Transformer-based model that considers document-level context for document reranking. Training a BERT model from scratch requires compute resource and larger data corpus. Approach like IR-BERT\cite{article5} ,  have used a combination of BM25 and sentence BERT for background linking task of news articles. This is similar to linking prior case with the queried legal case. We tried to follow the same approach for our task but were unsuccessful as pretrained BERT models did not generalise well on legal data unlike on news data. 
\section{Background}
The dataset used was provided by organisers \cite{article6} of AILA. The organisers of AILA collected all Indian Supreme Court case documents from 1952 to February 2018. Since the competition also had a task for statute retrieval, top 200 statutes were retrieved first, out of these 3 were removed as they were repeated.  Case documents that cited any of the selected 197 statutes were considered and 50 of these were picked randomly. The chronology of events that led to filing the case were extracted manually. Then, the prior cases cited in these 50 query were extracted and 2914 of them were provided as prior cases for the task’s corpus. We were also provided with ground truth values for each of the 50 queries. The distribution of relevant documents in the provided corpus is uneven and imbalanced. Some queries have as less as two relevant prior cases and some have about ten. Main goal is to retrieve most relevant prior cases among the 2914, for the given query. For our experiment the whole corpus of 2914 prior cases was used to find embeddings or for ranking them.\newline
\par We tried various methods for ranking these documents and found BM25 scores to be the best followed by cosine similarity on TF-IDF vectors of the documents. Since probabilistic model ranks the documents based on theoretical backing of probability and our dataset is unbalanced, it seemed to be a good fit unsupervised learning method for our project. There are other probabilistic models like BIM that perform reasonably well, but BIM is more efficient on short text like titles or abstracts. BM25 combines inverse document frequency, term frequency, document length and query term frequency to estimate relevance. BM25 does not need a lot of parameters and pays more attention to the rarer terms in a query by increasing their term weight while dampening the matching signal for words that occur too frequently in a document with a term saturation mechanism. Term weights are also normalized with respect to the document length.
\section{Pre-processing}
To make the document retrieval process more efficient we pre-processed the data. We have tried various permutations of standard pre-processing methods and applied only those pre-processing steps that improved performance of our approach. We have applied the same pre-processing on both prior cases and query documents.

\subsection{Lower Case}
Since our data is not very large converting all the words in corpus to lower case helps in maintaining consistency. Unlike programming languages, legal data is an NLP task where considering the case of each word is not critical. Converting all words into lower case has helped us in improving efficiency of our used approach.

\subsection{Noise removal} 
As part of pre-processing we have removed punctuation, special characters, numbers and words having length of less than two. Words with length less than two often are less significant.

\subsection{Stop-word Removal} 
Stop words are a set of commonly used words in a language. Examples of stop words in English are “a”, “the”, “is”, “are” and etc. Generally these word appears many times in the document and don’t contribute any information for the document. Stop-words list are can different for different document collection based on the purpose. We have removed the stop-words available in NLTK. The intuition behind using stop words is that, by removing low information words from text, we can focus on the important words instead. Since we are considering methods like BM25 and TF-IDF, removing stop words help in optimising the computations by preventing the stop-words from being analysed. 

\subsection{Stemming}
Stemming is the process of reducing inflection in words to their root form. The “root” in this case may not be a real root word, but just a canonical form of the original word. There are different algorithms for stemming, for our work we have used porter stemmer. Stemming is useful for dealing with sparsity issues as well as standardizing vocabulary. In case of our corpus we observed that stemming did not contribute to any performance improvement, this may be because matching all variations of a word like judge, judgement is necessary to bring up the most relevant documents.
\subsection{Tokenization}
To calculate TF-IDF scores or find embeddings using word2vec, splitting the document into token’s is helpful. We have done unigram tokenization of data. In general, tokens help in understanding the context.
\section{Methodology}
We are proposing  various unsupervised algorithms for identifying relevant prior cases. These algorithms are ranking the documents in the decreasing order of their similarity with the query and top N documents of the ranking are termed as relevant documents to the query Q.
\subsection{TF-IDF and Cosine similarity score}
This is one of the simplest and intuitive approach to solve the ad-hoc information retrieval task.\cite{article8} 
\begin{equation}
 TF-IDF_i = TF_i * \log \frac{N}{DF_i} 
\end{equation}
Here \emph{TF} (term frequency) is calculating the frequency of the term in the document and \emph{DF} (document frequency) is calculating the number of documents containing the term.
\emph{TF-IDF} is calculating the value for each term in the document. Terms with high tf-idf number imply a strong relationship with the document they appear in, suggesting that if that word were to appear in a query, the document could be an interest to the user. To compute the similarity score between every query Q and document D, we are first converting them into vectors, based on their TF-IDF values and then computing the cosine of the angle between the two vectors.
\begin{equation}
 similarity = cos(\theta) = \frac{A . B}{||A|| ||B||} = \frac{\sum_{i=1}^{n}A_i B_i}{\sqrt{\sum_{i=1}^{n} A_i ^2}\sqrt{\sum_{i=1}^{n} B_i ^2}}
\end{equation}

\subsection{Word Embeddings}
For finding word embeddings we experimented with word2vec, sent2vec and doc2vec methods. Word2Vec method has two variations the continuous bag of words (CBOW) and the skip-gram model. CBOW method takes the context of each word as the input and tries to predict the word corresponding to the context. Skip-gram does the opposite, and tries to predict several context words from a single input word. We have used pretrained genism word2vec model, we have tried both the CBOW and Skip-gram approach to find word embeddings. There was no significant change in the cosine similarity scores. We also used doc2vec for finding document embeddings. Doc2Vec is an unsupervised algorithm that learns fixed-length feature representations from variable-length pieces of texts, such as sentences, paragraphs, and documents. This algorithm represents each document by a dense vector that is trained to predict words in the document.Apart from these two methods, we also experimented with publicly available sentence2vec \cite{article7} embedding model which was trained on 53k Indian supreme court case documents. We observed that TF-IDF vectors outperformed the cosine similarity scores obtained by word embedding methods.
\subsection{Best Match 25}
BM25 is a probabilistic model that is developed by Stephen E Robertson, Karen Spark Jones, and others\cite{article10}. Here we are using another variant of BM25 given by Tortman \emph{et al.}\cite{article9} which has proven to be effective in a number of scenarios, and by independent authors\cite{article11}. 

\begin{equation}
rsv_q = \sum_{t\in q}\log\frac{N}{df_t}.\frac{(k_1 +1).tf_d}{k_1.(1-b+b.(\frac{L_d}{L_avg})) + tf_d}
\end{equation}

For a given query \emph{q}, the retrieval status value, \emph{rsv\textsubscript{q}}, is the sum of individual term, \emph{t} scores. \emph{N} is the number of documents in the collection, \emph{df\textsubscript{t}} is the number of the documents containing the term (the document frequency), \emph{tf\textsubscript{td}} is the number of times term \emph{t} occurs in the document \emph{d}. \emph{L\textsubscript{d}} is the length of the document (in terms) and \emph{L\textsubscript{avg}} is the mean of the document lengths. There are two tuning parameters, \emph{b}, and \emph{k\textsubscript{1}}. 

This ATIRE variant of BM25, is using Robertson-Walker \emph{IDF}[4], which tends to zero as \emph{df\textsubscript{t}} tends to \emph{N}. This ATIRE function always considers documents containing the query term to be more relevant than those that do not. We are using \emph{k\textsubscript{1}}=1.5, \emph{b}=0.75 and \emph{epsilon}=0.25 as the value of our tuning parameters. We are creating an indexed corpus using BM25 scores of every document. For every query we are searching this corpus and retrieving the top N documents as relevant documents.

\subsection{TF-IDF * BM25}
To retrieve the documents with the highest rank in both \emph{TF-IDF} and \emph{BM25}, we are multiplying the cosine similarity scores of a query-document pair with its BM25 score.\cite{coolie} We are again re-ranking the documents with this new score in sorted decreasing order to obtain top N relevant documents.
\subsection{Rake + TF-IDF and cosine similarity}
In this approach we are using Rapid automatic Keyword Extraction (RAKE)\cite{RAKE} algorithm, to extract top keywords from the data and query corpus. Only these keywords are then used to compute TF-IDF scores of every word and generate TF-IDF vectors for both query and document. Rankings are generated according to decreasing cosine similarity scores computed between query and document vectors and top N documents are retrieved as relevant.
\subsection{Sentence-BERT}
Both \emph{Tf-IDF} and \emph{BM25} are bag-of-words retrieval functions that rank a set of documents based on the query terms appearing in each document. In order to understand the context of the query Q and document D, it is important to take semantics of words into consideration. The \emph{state of the art} method to do so is BERT (Bidirectional Encoder Representations from Transformers). With lack of computing power and space to fine-tune BERT on our data set, we decided to use the Sentence-BERT model\cite{sbert}.

To compute the embeddings, first we are splitting our long documents into small paragraphs P\textsubscript{i} with less than 512 tokens to apply the SBERT. Then we are computing the embeddings for every query Q and splitted paragraph P\textsubscript{i}. The mean of cosine similarity between a query and a paragraph from the document is taken as the cosine similarity score between a query-document pair. This resulting score is used to rank the documents in decreasing order of their relevance with the query.
\section{Experiments}
A substantial number of ranking functions are discussed in Section 5. To analyse them better, we performed certain experiments on the data set by varying the intensity of pre-processing, using different variants of the same approach and multiplying certain results together to know their impact on the ranking results.

\subsection{BM25 Variants}
\begin{figure}[htp]
    \centering
    \includegraphics[width=8cm]{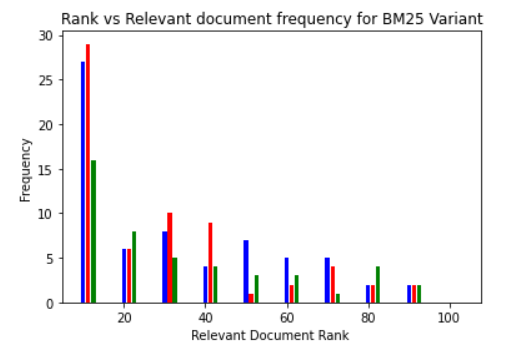}
    \caption{Rank vs Relevant document frequency for BM25}
    \label{fig:bm25variant}
\end{figure}

We tried different improvements suggested on BM25 to enhance our ranking function, and observed in figure 1 that among BM25Okapi (blue), BM25Plus (red) and BM25L (green), the frequency of relevant documents obtained within top 10 is more with BM25Plus.\cite{BM25Plus}

\subsection{Permutations of Pre-processing}
\begin{figure}[htp]
    \centering
    \includegraphics[width=8cm]{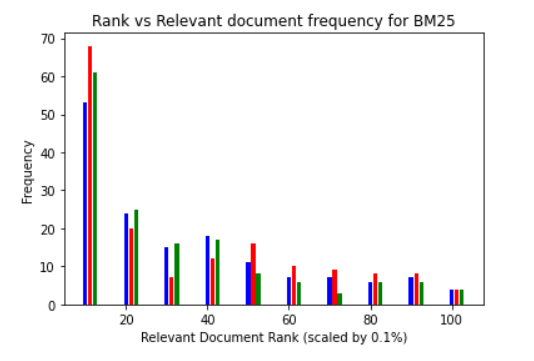}
    \caption{Rank vs Relevant document frequency for BM25}
    \label{fig:bm25variant}
\end{figure}

To measure the performance of BM25, we tried to create different BM25 indices from a data set with different permutations of pre-processing. The result of the number of relevant documents retrieved within different frequency classes can be seen in figure 2. Data set  with pre-processing using stop-word removal, lower casing every token and noise removal from the corpus (red) outperforms the other two which are no pre-processing (blue) and  pre-processing involving stemming using Porter stemmer, along with stop-word removal, lower casing every token and removing punctuation marks (green).
\subsection{Replacing text labels with NER tags}
\begin{figure}[htp]
    \centering
    \includegraphics[width=8cm]{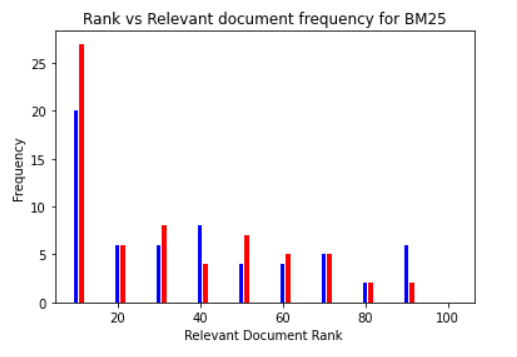}
    \caption{Rank vs Relevant document frequency for BM25}
    \label{fig:bm25variant}
\end{figure}

In our corpus we have many named entities for example name of the person, name of the organization, dates, numbers, etc. which we are replacing with their name entity recognition tags like Name, Organization, Cardinal etc. and observing the effect on the ranking of relevant documents in figure 3. We observed that NER tags replacing the text labels (blue) performs poor than the original text labels in the corpus (red) when we count the number of relevant documents retrieved within top 10.
\begin{table*}
  \caption{Results of Precedent retrieval for queries. All measures averaged over 50 test queries. Rows are sorted in decreasing order of P10 score (primary measure)}
  \label{tab:freq}
  \begin{tabular}{|c|c c c c|}
    \toprule
    Methods&P10&R10&F1 score& MRR\\
    \midrule
    BM25*(TF-IDF + Cosine similarity) &\textbf{0.06} &0.158& \textbf{0.075} & \textbf{0.26} \\
    TF-IDF + Cosine similarity & 0.054&0.152& 0.068 & 0.25 \\
    BM25 & 0.0539 &0.152& 0.070 & 0.213 \\
    Rake + TF-IDF + Cosine similarity & 0.048 &0.139& 0.0616 & 0.1824 \\
    SBERT + Cosine similarity & 0.01& 0.038 &0.0137&0.067\\
    Sent2vec & 0.006& \textbf{0.233}&0.0072&0.0225\\
    
  \bottomrule
\end{tabular}
\end{table*}

\subsection{Common words between query-document pair }
\begin{figure}[htp]
    \centering
    \includegraphics[width=8cm]{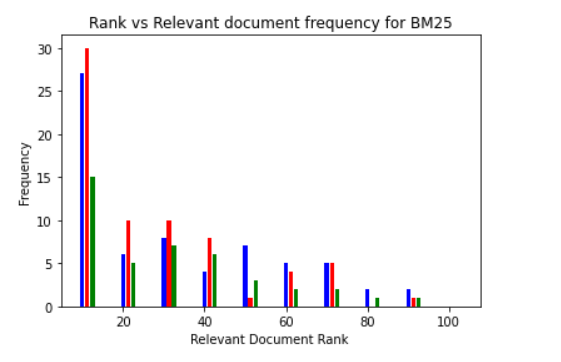}
    \caption{Rank vs Relevant document frequency for BM25}
    \label{fig:bm25variant}
\end{figure}

For this experiment, we are calculating the number of common words between every query-document pair and multiplying this number with their BM25 scores. The main intuition behind this experiment is more the number of common words, more relevant the document is to the query. We observed in figure 4 that BM25 (blue) and BM25*TF-IDF (red) performs better than number of common words * BM25 scores (green). Though we observed the high count of common words between query and document, its relevance with the query cannot be judged based on that. 

\begin{table}
  \caption{Comparison of P10 and MRR between proposed method and AILA 2020 submissions. All measures are averaged over 10 test queries.}
  \label{tab:freq2}
  \begin{tabular}{|c|c c|}
    \toprule
    Methods&P10 & MRR\\
    \midrule
    Language Model, Dirichlet smoothing&\textbf{0.1}&0.2041\\
    Topic Embedding&\textbf{0.1}&0.1586\\
    Terrier 4.2 KL Divergence model&0.08&0.238\\
    \midrule
    BM25*(TF-IDF + Cosine similarity) &\textbf{0.13}&\textbf{0.26} \\
  \bottomrule
\end{tabular}
\end{table}

\begin{table}
  \caption{ Precision at 1,3,5 and 10 averaged over 50 queries for our best performing method}
  \label{tab:freq3}
  \begin{tabular}{|c|c|}
    \toprule
    Metric& BM25*(TF-IDF + Cosine similarity)\\
    \midrule
    Precision@1&0.11\\
    Precision@3&0.108\\
    Precision@5&0.02\\
    Precision@10&0.06\\
  \bottomrule
\end{tabular}
\end{table}
\section{Results}
The primary metric used to evaluate the identification of relevant prior cases task is Precision at 10 (P10), Recall at 10 (R10), F1 score and Mean reciprocal rank (MRR).
\\
\textbf{P10:} Number of relevant documents in the top 10 ranked
results, averaged over all queries. Since each query contains 10 precedents on average, we report this score for the runs.
\\
\textbf{R10:} Number of relevant documents out of total relevant documents in the top 10 ranked results, averaged over all queries.
\\
\textbf{F1 score:} Harmonic mean of P10 and R10 results.
\\
\textbf{MRR:} inverse of the position of the \emph{first} relevant document in the ranked list of a query, averaged over all queries.

In our first experiment, we list P10,  Table~\ref{tab:freq} it is observed that \emph{BM25*(TF-IDF + Cosine similarity)} performed best on our primary measure P10, F1 score and MRR. Word embeddings like Sent2vec is giving us better recall as compared to other methods.s. It is interesting to note that using sBERT for finding semantic similarity between query and documents harms the performance given by BM25 and TF-IDF.

In our next set of experiments in Table~\ref{tab:freq2} we compare our best performing model \emph{BM25*(TF-IDF + Cosine similarity)} with several other approaches proposed at FIRE 2020 for the same task. 

In Table~\ref{tab:freq3} we computed Precision at 1, 3, 5 and 10 for our best performing approach.
\section{Conclusion and Future Work}
In this paper, we described various methods to relevant prior cases for a given query. The approaches that we used vary from using word embeddings like word2vec,doc2vec and sentence vector, followed by cosine similarity to finding TF-IDF vector of the documents followed by cosine similarity. We have also used probabilistic model like BM25 and pretrained BERT model like sBERT. We have provided a comparative analysis of how each of these methods performed for the given document retrieval task. Through the final evaluation results, it can be seen that the TF-IDF-based cosine similarity multiplied with BM25 gives the best results. We also briefed about the pre-processing steps that we have used. As a future scope of work deep learning based supervised approaches can be tried to solve the problem, but it may require more data as most deep-learning approaches are data-hungry. We can also train a BERT model from scratch to consider the semantic meaning of prior cases and query words, but data is the main bottleneck. We should put our focus on preparing more judicial domain annotated data for these problems. 
\begin{acks}
This project was given to us as a coursework for CS6730. We are grateful to our professor Dr. Maunendra Sankar Desarkar and our mentor Mr. Samujjwal Ghosh for providing us this opportunity to work on this challenging task and guiding us throughout the project.
\end{acks}

\bibliographystyle{ACM-Reference-Format}
\bibliography{sample-base}


\begin{thebibliography}{15}

%%% the Web address.

\ifx \showCODEN    \undefined \def \showCODEN     #1{\unskip}     \fi
\ifx \showDOI      \undefined \def \showDOI       #1{#1}\fi
\ifx \showISBNx    \undefined \def \showISBNx     #1{\unskip}     \fi
\ifx \showISBNxiii \undefined \def \showISBNxiii  #1{\unskip}     \fi
\ifx \showISSN     \undefined \def \showISSN      #1{\unskip}     \fi
\ifx \showLCCN     \undefined \def \showLCCN      #1{\unskip}     \fi
\ifx \shownote     \undefined \def \shownote      #1{#1}          \fi
\ifx \showarticletitle \undefined \def \showarticletitle #1{#1}   \fi
\ifx \showURL      \undefined \def \showURL       {\relax}        \fi
% invisible to TeX
\providecommand\bibfield[2]{#2}
\providecommand\bibinfo[2]{#2}
\providecommand\natexlab[1]{#1}
\providecommand\showeprint[2][]{arXiv:#2}

\bibitem[\protect\citeauthoryear{Baban~Gain and Ekbal}{Baban~Gain and
  Ekbal}{[n.d.]}]%
        {article3}
\bibfield{author}{\bibinfo{person}{Tanik~Saikh Baban~Gain,
  Dibyanayan~Bandyopadhyay} {and} \bibinfo{person}{Asif Ekbal}.}
  \bibinfo{year}{[n.d.]}\natexlab{}.
\newblock \showarticletitle{IITP in COLIEE@ICAIL 2019: Legal Information
  Retrieval usingBM25 and BERT}.
\newblock  (\bibinfo{year}{[n.\,d.]}).
\newblock
\urldef\tempurl%
\url{https://www.researchgate.net/publication/334112555_IITP_in_COLIEEICAIL_2019_Legal_Information_Retrieval_using_BM25_and_BERT}
\showURL{%
\tempurl}


\bibitem[\protect\citeauthoryear{Bhattacharya, Ghosh, Ghosh, Pal, Mehta,
  Bhattacharya, and Majumder}{Bhattacharya et~al\mbox{.}}{2019}]%
        {article7}
\bibfield{author}{\bibinfo{person}{Paheli Bhattacharya},
  \bibinfo{person}{Kripabandhu Ghosh}, \bibinfo{person}{Saptarshi Ghosh},
  \bibinfo{person}{Arindam Pal}, \bibinfo{person}{Parth Mehta},
  \bibinfo{person}{Arnab Bhattacharya}, {and} \bibinfo{person}{Prasenjit
  Majumder}.} \bibinfo{year}{2019}\natexlab{}.
\newblock \showarticletitle{Overview of the {FIRE} 2019 {AILA} Track:
  Artificial Intelligence for Legal Assistance}.
\newblock   \bibinfo{volume}{2517} (\bibinfo{year}{2019}),
  \bibinfo{pages}{1--12}.
\newblock
\urldef\tempurl%
\url{http://ceur-ws.org/Vol-2517/T1-1.pdf}
\showURL{%
\tempurl}


\bibitem[\protect\citeauthoryear{Deshmukh and Sethi}{Deshmukh and
  Sethi}{[n.d.]}]%
        {article5}
\bibfield{author}{\bibinfo{person}{Anup~Anand Deshmukh} {and}
  \bibinfo{person}{Udhav Sethi}.} \bibinfo{year}{[n.d.]}\natexlab{}.
\newblock \showarticletitle{IR-BERT: Leveraging BERT for Semantic Search in
  Background Linking for News Articles}.
\newblock  (\bibinfo{year}{[n.\,d.]}).
\newblock
\urldef\tempurl%
\url{https://arxiv.org/pdf/2007.12603.pdf}
\showURL{%
\tempurl}


\bibitem[\protect\citeauthoryear{et~al.}{et~al.}{[n.d.]}]%
        {article4}
\bibfield{author}{\bibinfo{person}{Canjia~Li1 et al.}}
  \bibinfo{year}{[n.d.]}\natexlab{}.
\newblock \showarticletitle{PARADE: Passage Representation Aggregation for
  Document Reranking}.
\newblock  (\bibinfo{year}{[n.\,d.]}).
\newblock
\urldef\tempurl%
\url{https://arxiv.org/pdf/2008.09093v1.pdf}
\showURL{%
\tempurl}


\bibitem[\protect\citeauthoryear{Jangid, Singhania, Rajasthan, Vishwakarma,
  Scholar, Singhania, Rajasthan, Lakhtaria, Singhania, and Rajasthan}{Jangid
  et~al\mbox{.}}{[n.d.]}]%
        {article8}
\bibfield{author}{\bibinfo{person}{Chandra~Shekhar Jangid},
  \bibinfo{person}{Sir~Padampat Singhania}, \bibinfo{person}{Udaipur
  Rajasthan}, \bibinfo{person}{Santosh~K Vishwakarma}, \bibinfo{person}{Ph.~D.
  Scholar}, \bibinfo{person}{Sir~Padampat Singhania}, \bibinfo{person}{Udaipur
  Rajasthan}, \bibinfo{person}{Kamaljit~I Lakhtaria},
  \bibinfo{person}{Sir~Padampat Singhania}, {and} \bibinfo{person}{Udaipur
  Rajasthan}.} \bibinfo{year}{[n.d.]}\natexlab{}.
\newblock \bibinfo{title}{Ad-hoc Retrieval on FIRE Data Set with TF-IDF and
  Probabilistic Models}.
\newblock
\newblock


\bibitem[\protect\citeauthoryear{Lv and Zhai}{Lv and Zhai}{2011}]%
        {BM25Plus}
\bibfield{author}{\bibinfo{person}{Yuanhua Lv} {and}
  \bibinfo{person}{ChengXiang Zhai}.} \bibinfo{year}{2011}\natexlab{}.
\newblock \showarticletitle{Lower-Bounding Term Frequency Normalization}. In
  \bibinfo{booktitle}{\emph{Proceedings of the 20th ACM International
  Conference on Information and Knowledge Management}} (Glasgow, Scotland, UK)
  \emph{(\bibinfo{series}{CIKM '11})}. \bibinfo{publisher}{Association for
  Computing Machinery}, \bibinfo{address}{New York, NY, USA},
  \bibinfo{pages}{7–16}.
\newblock
\showISBNx{9781450307178}
\urldef\tempurl%
\url{https://doi.org/10.1145/2063576.2063584}
\showDOI{\tempurl}


\bibitem[\protect\citeauthoryear{Moemedi~Lefoane and
  Narasimham}{Moemedi~Lefoane and Narasimham}{[n.d.]}]%
        {article1}
\bibfield{author}{\bibinfo{person}{Goaletsa~Rammidi Moemedi~Lefoane,
  Tshepho~Koboyatshwene} {and} \bibinfo{person}{V.Lakshmi Narasimham}.}
  \bibinfo{year}{[n.d.]}\natexlab{}.
\newblock \showarticletitle{Legal Statutes Retrieval: A Comparative Approach on
  Performance of Title and Statutes Descriptive Text}.
\newblock  (\bibinfo{year}{[n.\,d.]}).
\newblock
\urldef\tempurl%
\url{http://ceur-ws.org/Vol-2517/T1-9.pdf}
\showURL{%
\tempurl}


\bibitem[\protect\citeauthoryear{Paheli~Bhattacharya1 and
  Majumder}{Paheli~Bhattacharya1 and Majumder}{[n.d.]}]%
        {article6}
\bibfield{author}{\bibinfo{person}{Saptarshi Ghosh Arindam Pal Parth Mehta
  Arnab~Bhattacharya Paheli~Bhattacharya1, Kripabandhu~Ghosh} {and}
  \bibinfo{person}{Prasenjit Majumder}.} \bibinfo{year}{[n.d.]}\natexlab{}.
\newblock \showarticletitle{Overview of the FIRE 2019 AILA Track: Artificial
  Intelligence for Legal Assistance}.
\newblock  (\bibinfo{year}{[n.\,d.]}).
\newblock
\urldef\tempurl%
\url{http://ceur-ws.org/Vol-2517/T1-1.pdf}
\showURL{%
\tempurl}


\bibitem[\protect\citeauthoryear{Petri, Moffat, and Culpepper}{Petri
  et~al\mbox{.}}{2014}]%
        {article11}
\bibfield{author}{\bibinfo{person}{Matthias Petri}, \bibinfo{person}{Alistair
  Moffat}, {and} \bibinfo{person}{J.~Shane Culpepper}.}
  \bibinfo{year}{2014}\natexlab{}.
\newblock \showarticletitle{Score-Safe Term-Dependency Processing with Hybrid
  Indexes}. In \bibinfo{booktitle}{\emph{Proceedings of the 37th International
  ACM SIGIR Conference on Research and Development in Information Retrieval}}
  (Gold Coast, Queensland, Australia) \emph{(\bibinfo{series}{SIGIR '14})}.
  \bibinfo{publisher}{Association for Computing Machinery},
  \bibinfo{address}{New York, NY, USA}, \bibinfo{pages}{899–902}.
\newblock
\showISBNx{9781450322577}
\urldef\tempurl%
\url{https://doi.org/10.1145/2600428.2609469}
\showDOI{\tempurl}


\bibitem[\protect\citeauthoryear{Rabelo, Kim, Goebel, Yoshioka, Kano, and
  Satoh}{Rabelo et~al\mbox{.}}{2020}]%
        {coolie}
\bibfield{author}{\bibinfo{person}{Juliano Rabelo}, \bibinfo{person}{Mi-Young
  Kim}, \bibinfo{person}{Randy Goebel}, \bibinfo{person}{Masaharu Yoshioka},
  \bibinfo{person}{Yoshinobu Kano}, {and} \bibinfo{person}{Ken Satoh}.}
  \bibinfo{year}{2020}\natexlab{}.
\newblock \showarticletitle{A Summary of the COLIEE 2019 Competition}. In
  \bibinfo{booktitle}{\emph{New Frontiers in Artificial Intelligence}},
  \bibfield{editor}{\bibinfo{person}{Maki Sakamoto}, \bibinfo{person}{Naoaki
  Okazaki}, \bibinfo{person}{Koji Mineshima}, {and} \bibinfo{person}{Ken
  Satoh}} (Eds.). \bibinfo{publisher}{Springer International Publishing},
  \bibinfo{address}{Cham}.
\newblock
\showISBNx{978-3-030-58790-1}


\bibitem[\protect\citeauthoryear{Reimers and Gurevych}{Reimers and
  Gurevych}{2019}]%
        {sbert}
\bibfield{author}{\bibinfo{person}{Nils Reimers} {and} \bibinfo{person}{Iryna
  Gurevych}.} \bibinfo{year}{2019}\natexlab{}.
\newblock \showarticletitle{Sentence-BERT: Sentence Embeddings using Siamese
  BERT-Networks}. In \bibinfo{booktitle}{\emph{EMNLP/IJCNLP}}.
\newblock


\bibitem[\protect\citeauthoryear{Renjit1 and Idicula2}{Renjit1 and
  Idicula2}{[n.d.]}]%
        {article2}
\bibfield{author}{\bibinfo{person}{Sara Renjit1} {and}
  \bibinfo{person}{Sumam~Mary Idicula2}.} \bibinfo{year}{[n.d.]}\natexlab{}.
\newblock \showarticletitle{CUSAT NLP@AILA-FIRE2019: Similarity in Legal Texts
  using Document Level Embeddings}.
\newblock  (\bibinfo{year}{[n.\,d.]}).
\newblock
\urldef\tempurl%
\url{http://ceur-ws.org/Vol-2517/T1-4.pdf}
\showURL{%
\tempurl}


\bibitem[\protect\citeauthoryear{Robertson, Walker, Beaulieu, Gatford, and
  Payne}{Robertson et~al\mbox{.}}{1996}]%
        {article10}
\bibfield{author}{\bibinfo{person}{S.E. Robertson}, \bibinfo{person}{S.
  Walker}, \bibinfo{person}{M.M. Beaulieu}, \bibinfo{person}{M. Gatford}, {and}
  \bibinfo{person}{A. Payne}.} \bibinfo{year}{1996}\natexlab{}.
\newblock \showarticletitle{Okapi at TREC-4}. In \bibinfo{booktitle}{\emph{In
  Proceedings of the 4th Text REtrieval Conference (TREC-4}}.
  \bibinfo{pages}{73--96}.
\newblock


\bibitem[\protect\citeauthoryear{Rose, Engel, Cramer, and Cowley}{Rose
  et~al\mbox{.}}{2010}]%
        {RAKE}
\bibfield{author}{\bibinfo{person}{S. Rose}, \bibinfo{person}{D. Engel},
  \bibinfo{person}{Nick Cramer}, {and} \bibinfo{person}{W. Cowley}.}
  \bibinfo{year}{2010}\natexlab{}.
\newblock \showarticletitle{1 Automatic keyword extraction from individual
  documents}.
\newblock


\bibitem[\protect\citeauthoryear{Trotman, Puurula, and Burgess}{Trotman
  et~al\mbox{.}}{2014}]%
        {article9}
\bibfield{author}{\bibinfo{person}{A. Trotman}, \bibinfo{person}{Antti
  Puurula}, {and} \bibinfo{person}{B. Burgess}.}
  \bibinfo{year}{2014}\natexlab{}.
\newblock \showarticletitle{Improvements to BM25 and Language Models Examined}.
  In \bibinfo{booktitle}{\emph{ADCS '14}}.
\newblock


\end{thebibliography}
\end{document}